\documentclass[conference]{IEEEtran}
\IEEEoverridecommandlockouts
\usepackage{cite}
\usepackage{amsmath,amssymb,amsfonts}
\usepackage{algorithmic}
\usepackage{graphicx}
\usepackage{textcomp}
\usepackage{xcolor}
\usepackage{float}
\usepackage[colorinlistoftodos]{todonotes}
\def\BibTeX{{\rm B\kern-.05em{\sc i\kern-.025em b}\kern-.08em
    T\kern-.1667em\lower.7ex\hbox{E}\kern-.125emX}}
\begin{document}
\renewcommand\citepunct{, }
\title{Decision SincNet: Neurocognitive models of decision making that predict cognitive processes from neural signals}


\author{\IEEEauthorblockN{Qinhua Jenny Sun}
\IEEEauthorblockA{\textit{Department of Cognitive Sciences} \\
\textit{University of California, Irvine}\\
Irvine, California, USA \\
qinhuas@uci.edu}
\and
\IEEEauthorblockN{Khuong Vo}
\IEEEauthorblockA{Department of Computer Science \\
\textit{University of California, Irvine}\\
Irvine, California, USA \\
khuongav@uci.edu}

\and
\IEEEauthorblockN{Kitty Lui}
\IEEEauthorblockA{Department of Psychiatry
 \\
\textit{University of California, San Diego}\\
La Jolla, California, USA \\
klui@health.ucsd.edu }

\and
\IEEEauthorblockN{Michael Nunez}
\IEEEauthorblockA{Psychological Methods Group\\
\textit{University of Amsterdam}\\
Amsterdam, Netherlands \\
m.d.nunez@uva.nl}

\and
\IEEEauthorblockN{Joachim Vandekerckhove}
\IEEEauthorblockA{Department of Cognitive Sciences \\
\textit{University of California, Irvine}\\
Irvine, California, USA \\
joachim@uci.edu}

\and
\IEEEauthorblockN{Ramesh Srinivasan}
\IEEEauthorblockA{Department of Cognitive Sciences\\
\textit{University of California, Irvine}\\
Irvine, California, USA  \\
srinivar@uci.edu }
}

\maketitle

\begin{abstract}
Human decision making behavior is observed with choice-response time data  during psychological experiments. Drift-diffusion models of this data consist of a Wiener first-passage time (WFPT) distribution and are described by cognitive parameters: drift rate, boundary separation, and starting point. These estimated parameters are of interest to neuroscientists as they can be mapped to features of cognitive processes of decision making (such as speed, caution, and bias) and related to brain activity. The observed patterns of RT also reflect the variability of cognitive processes from trial to trial mediated by neural dynamics. We adapted a SincNet-based shallow neural network architecture to fit the Drift-Diffusion model using EEG signals on every experimental trial. The model consists of a SincNet layer, a depthwise spatial convolution layer, and two separate fully connected layers that predict drift rate and boundary for each trial in-parallel. The SincNet layer parametrized the kernels in order to directly learn the low and high cutoff frequencies of bandpass filters that are applied to the EEG data to predict drift and boundary parameters. During training, model parameters were updated by minimizing the negative log likelihood function of WFPT distribution given trial RT. We developed separate decision SincNet models for each participant performing a two-alternative forced-choice task. Our results showed that single-trial estimates of drift and boundary performed better at predicting RTs than the median estimates in both training and test data sets, suggesting that our model can successfully use EEG features to estimate meaningful single-trial Diffusion model parameters. Furthermore, the shallow SincNet architecture identified time windows of information processing related to evidence accumulation and caution and the EEG frequency bands that reflect these processes within each participant. 
\end{abstract}

\begin{IEEEkeywords}
EEG, Diffusion Model, Neural Network
\end{IEEEkeywords}

\section{Introduction}

Perceptual decision making is a crucial part of cognition. Humans must rapidly translate sensory information into behavioral responses in order to achieve their goals, e.g., stop at a red traffic light. In the field of mathematical psychology and computational neuroscience, much effort has been dedicated to develop computational models that describe the mechanism of perceptual decision-making. The Drift-Diffusion Model (DDM) is the most widely used model \cite{ratcliff2008diffusion} to explain choice and response time data in perceptual decision making tasks, by assuming that decisions are made through sequential sampling and integration of sensory information\cite{nunez2017attention}.

\begin{figure}[h]
\centerline{\includegraphics[scale=0.35]{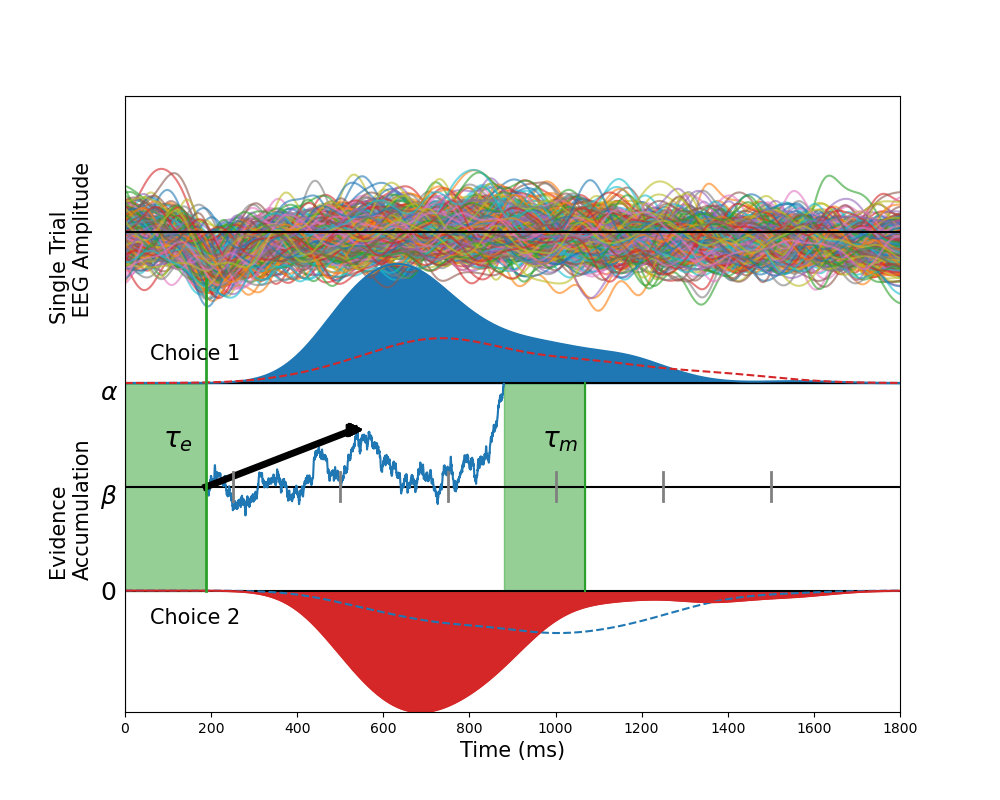}}

\caption{Demonstration of Drift Diffusion Model during a  two-choice decision making task with Non-deicison time indicated in green. For a given trial, after visual encoding time ($\tau_e$), a DV begins evidence accumulation process and will hit either the upper of the lower bound. Mean rate of evidence accumulation of is indicated by the vector in black. Blue curve indicates the probability density function (pdf) of RTs when Choice 1 is correctly executed, and red curve indicates the pdf of RTs when Choice 2 is correctly. An error is made when the DV drifts to the wrong bound due to noise. Dotted curve indicates the pdf of RTs for incorrect trials. The axis above shows the EEG data for each trial after using SVD to maximize N200 signals, which can be used to track onset of evidence accumulation.}

\label{fig1}
\end{figure}

Fig.~\ref{fig1} depicts the Drift-Diffusion Model (DDM), where the x-axis is the time from viewing of a stimulus. After some processing time for neural encoding of the sensory stimulus, a Decision Variable (DV) begins a random walk process from starting point $z$ between two decision boundaries, which represent the two choice options. The DV is updated by gradually accumulating samples drawn from noisy sensory evidence in favor of one of the two outcomes, until it reaches either the upper or the lower bound. Subsequently, the motor system then executes a response after the DV reaches one of two boundaries. 

The average increase of change in a unit interval during evidence accumulation is parameterized as the drift rate $\delta$, and the instantaneous variance in the rate of change is parameterized as diffusion coefficient $\varsigma$.  This variance simply scales the model and is fixed to 1. Boundary separation $\alpha$ describes the distance between two choices. The parameter $\beta$ encodes the starting position of evidence accumulation, which reflects the bias towards one of the two choices. When $\beta$ is 0.5, the starting position is in the middle between two boundaries and thus the start of evidence accumulation is unbiased between the two choices. Visual encoding time before evidence accumulation and motor execution after evidence accumulation could be expressed as $\tau_e$ and $\tau_m$ \cite{nunez2019latency} respectively, but only the sum of both processes, non-decision time $\tau$, can  be observed with behavior alone. The model is also referred to as Wiener Diffusion Model, because an unbiased, continuous evidence accumulation can be expressed as follow\cite{bogacz2006physics}: 
\begin{equation}
\begin{array}{r}
dX_t = \delta dt + \varsigma dW_t, \; x(0).
\end{array}
\end{equation} 
$X_t$ denotes the diffusion state, $dX_t$ denotes the change in $X$ over a small time interval $dt$, and $W_t$ is the Wiener Process. Evidence accumulated continuously will result in the distribution of boundary cross times,  described as the Wiener first-passage time
(WFPT) distribution:
$ \mathcal{T} \sim Wiener(\alpha,\beta,\tau,\delta). $

One of the most critical aspects of the DDM is that different parameters can empirically represent underlying cognitive components of the decision making process. During decision making experiments, separate manipulation of processing speed, caution, bias, and motor execution could be reflected in changes in the corresponding parameters $\delta$, $\alpha$, $\beta$, and $\tau$ \cite{voss2004interpreting}.  Thus, there has been considerable interest in linking the DDM parameters to brain activity, and more recently, interest in developing neurocognitive models that incorporate brain activity directly into the diffusion model framework to obtain novel insights into decision making, in particular, to the variability in decision making across observations.

\section{Relevant Work}

Simulation studies with the DDM have demonstrated the importance of trial-to-trial variability in drift rate, starting point, and boundary to empirically model two-choice response time data  \cite{ratcliff1998modeling}. Allowing drift rate and starting point to vary from trial to trial can uniquely capture properties of response time distribution for correct and incorrect choices. Variability in caution reflect adjustments in speed-accuracy trade-off. 

Although variability in cognitive processes in decision is well-known, choice-response time does not provide enough data to estimate model parameters at a trial level. Often models with model-intrinsic trial-to-trial variability, such as assuming a normal distribution for trial-to-trial drift rates are fit to data. Fitting these models provide estimates of summary parameters across trials (e.g. mean and standard deviation of the drift rates across trials). One method to estimate DDMs while allowing intrinsic trial-to-trial variability of parameters is to use Bayesian Hierarchical Diffusion Models \cite{vandekerckhove2011hierarchicalvandekerckhove2011hierarchical}. The Bayesian modeling approach is used to compute
posterior distribution of model parameters, and employs Monte Carlo Markov Chain (MCMC) algorithm to generate samples from the  posterior distribution until they converge. 

 Accumulation-to-bound patterns have also been found in different areas of the brain\cite{roitman2002response, OConnell2018}. There has been a growing interest to incorporate neural data along with behavioral data to build \emph{neurocognitive} models of decision making \cite{nunez2017attention,nunez2019latency,turner2013bayesian, Turner2017}. Previous work has shown that incorporating single-trial EEG measures of attention into the HDDM yields better out-of-sample predictions on accuracy and reaction time distributions relative to using behavioral data alone\cite{nunez2017attention}. Model parameters ($\delta$, $\tau$, $\varsigma$) on each trial were assumed to be a linear combination of single-trial EEG. The EEG measures were derived using known stimulus-locked EEG signals, i.e., event-related potentials (ERPs) estimated by averaging the trials. More specifically, to augment the signal-to-noise (SNR) ratio of single-trial EEG, singular value decomposition (SVD) was used to find channel weights that maximally explain the variance of specific evoked potentials (N200, P200). These weights were applied to single-trial EEG to obtain latency and amplitude measures of the N200/P200 per trial as regressors onto HDDM parameters. This successful line of work has suggested that trial-level neural dynamics account for some of the trial-to-trial variability in the HDDM model parameters. However, while using MCMC alone can robustly estimate posterior distributions from which the trial parameters are drawn, it does not have the resolution to directly link trial-level neural representation to trial-level parameters estimated. Moreover, this approach was limited to using trial averaged ERP signals to provide a template of hypothetical signals linked to DDM parameters.  
 
 The current research aims to use a neural network to estimate drift rate $\delta$ and boundary $\alpha$ parameters of the DDM on single trials directly from the raw EEG data. Machine learning approaches
 such as filter bank common spatial patterns (FBCSP) have been widely used to extract EEG features \cite{ang2012filter}, but it has the disadvantage of requiring artificially-selected frequency bands. Convolutional Neural Networks (CNNs) have shown promising results on decoding brain activities \cite{amin2019deep, zhang2020investigation} using raw EEG data, but filters and feature maps obtained can be hard to interpret. SincNet is a recently proposed deep learning neural network used to process raw time series data such as speech and EEG data\cite{ravanelli2018speaker}. The key feature of SincNet is that each kernel from the first layer of CNN is parameterized as a sinc function and acts as a band-pass filter that could be applied to the time series. Two cut-off frequencies are the only trainable parameters needed. Therefore, SincNet is advantageous because fewer parameters are needed and the filter parameters are themselves optimized by the training data. The model structure has successfully been used on different EEG decoding tasks since its inception\cite{borra2020interpretable, mayor2021interpretable}.

We developed a Decision SincNet model that can learn the windows of time and frequency bands in the EEG that are useful to predict trial level DDM parameters. This Decision SincNet model is scientifically significant for the following reasons.

\begin{enumerate}
\item By using the WFPT likelihood as a loss function, we can use gradient descent to learn an end-to-end model to fit two DDM parameters (drift rate and boundary) from raw EEG data, and apply it to predict cognitive parameters in new unseen brain data. 

\item By using neural network models, we can get trial-level predictions of parameters

\item By applying interpretation techniques to the SincNet layer and depthwise layer, we can automatically learn relevant neural dynamics (e.g., time windows and frequency bands) that are critical to modeling the evidence accumulation process.
\end{enumerate}


\section{Methods}

\subsection{Behavioral and EEG dataset}
We used behavioral and EEG data collected while participants performed a two-alternative forced-choice task where they discriminated whether a Gabor patch presented with added dynamic noise is higher or lower spatial frequency. Task difficulty was manipulated by the difference in spatial frequency (perceptual evidence) between the two choices in order to manipulate the evidence available to make the discrimination. Participants performed the task in blocks of trials at 3 levels of difficulty (spatial frequency difference). Each subject performed 360 trials, while 128 channels of EEG and behavioral data were recorded. Independent Component Analysis (ICA) based artifact rejection method was used on EEG data to remove eyeblinks, electrical noise, and muscle artifact. EEG data were bandpass filtered to 1 to 50 Hz, and then downsampled from 1000 Hz to 500 Hz prior to data analysis. 

\subsection{Wiener first-passage time model of response time}



For each trial, the likelihood of Wiener First-Passage Time (WFPT) was calculated with RT ($t$) by using a small-time approximation

\begin{equation}
\begin{array}{r}
\textrm{Wiener}(t \mid \delta, \alpha, \beta, \tau)= \frac{1}{\alpha^{2}}\exp\left(-\delta \alpha \beta-\frac{\delta^{2} t}{2}\right) \\
\times \frac{1}{\sqrt{2 \pi(t-\tau)^{3}}} \sum\limits_{\-[(m-1) / 2]}^{[(m-1) / 2}(\alpha+2 m) \\
\times \exp \left(-\frac{(\beta+2 m)^{2}}{2(t-\tau)}\right)
\end{array}
\end{equation}

where $m$ is the number of expansion terms required for approximation of the likelihood function. We fix $m$ to be 10, as it is sufficient for the approximation for modeling data where $t <2 $\cite{navarro2009fast}. $\beta$ is set to be 0.5, so that the starting point is always unbiased at $z$ = $\beta \alpha$. $\tau$ is non-decision time, which is set to be $0.93\cdot RT_{min}$ for each subject, approximating Bayesian MCMC modeling results \cite{nunez2019latency}.

Given a dataset $\mathcal{D}:=\left\{\left(\boldsymbol{x}_{1}, y_{1}\right), \ldots,\left( \boldsymbol{x}_{N}, y_{N}\right)\right\}$ consisting of $N$ trials $\boldsymbol{x}_{n} \in$ $\mathbb{R}^{C \times D}$, i.e., EEG signals of $C$ channels by $N$ time samples, and corresponding observed response times $t_{n} \in \mathbb{R}, n=1, \ldots, N$. The likelihood factorizes according to
\begin{equation}
\begin{aligned}
p(\mathcal{T} \mid \mathcal{X}, \boldsymbol{\theta}) &=p\left(t_{1}, \ldots, t_{N} \mid \boldsymbol{x}_{1}, \ldots, \boldsymbol{x}_{N}, \boldsymbol{\theta}\right) \\
&=\prod_{n=1}^{N} p\left(t_{n} \mid \boldsymbol{x}_{n}, \boldsymbol{\theta}\right)\\ &=\prod_{n=1}^{N} \textrm{Wiener}\left(t_{n} \mid \delta(\boldsymbol{x}_{n}, \boldsymbol{\theta}), \alpha(\boldsymbol{x}_{n}, \boldsymbol{\theta}), \beta, \tau\right)
\end{aligned}
\end{equation}
where we defined $\mathcal{X}:=\left\{\boldsymbol{x}_{1}, \ldots, \boldsymbol{x}_{N}\right\}$ and $\mathcal{T}:=\left\{t_{1}, \ldots, t_{N}\right\}$ as the sets of inputs and corresponding targets, respectively. Both the drift rate $\delta$ and the boundary $\alpha$ are the functions of $\boldsymbol{x}$ parameterized by the a deep neural network $\boldsymbol{\theta}$.

To find the optimal parameters $\boldsymbol{\theta}^{\star}$ of the non-linear regression problem, we minimize the negative log-likelihood

\begin{equation}
\begin{aligned}
\boldsymbol{\theta}_{\star}&=\arg \min _{\boldsymbol{\theta}} \mathcal{L}(\theta)
:=\arg \min _{\boldsymbol{\boldsymbol{\theta}}} -\log p(\mathcal{T} \mid \mathcal{X}, \boldsymbol{\theta})\\&=\arg \min _{\boldsymbol{\theta}} -\sum_{n=1}^{N} \log  \textrm{Wiener}\left(t_{n} \mid \delta(\boldsymbol{x}_{n}, \boldsymbol{\theta}), \bar \alpha(\boldsymbol{x}_{n}, \boldsymbol{\theta}), \beta, \tau\right)
\end{aligned}
 \end{equation}

where $\bar \alpha$ is the average boundary within a training batch, in order to avoid overfitting of the boundary parameter. Since the gain of varying the boundary parameter outweighs the gain of varying the drift parameter given the nature of the WFPT likelihood function, we use this method to update trial-level boundary while considering the fit within a batch. 





\begin{table*}[t]
\centering
\caption{Architecture of the Decision SincNet Model}
\label{tab:net-table}
\begin{tabular}{lllllll}
\textbf{Block} & \textbf{Layer} & \textbf{\# filters} & \textbf{size} & \textbf{\# params} & \textbf{Output} & \textbf{Activation} \\ \hline
1 & Input              &    &          &      & (98, 500)     &         \\
  & BatchNorm          &    &          & 2    & (98, 500)     &         \\
  & SincConv2D         & 32 & (1, 131) & 64   & (32, 98, 370) &         \\ \hline
2 & BatchNorm          &    &          & 64   & (32, 98, 370) &         \\
  & SeparableConv2D    & 64 & (98, 1)  & 6272 & (64, 1, 370)  & ReLU    \\ \hline
3 & BatchNorm          &    &          & 128  & (64, 1, 370)  &         \\
  & AvgPool2D          &    & (1, 45)  &      & (64, 1, 8)    &         \\
  & Dropout            &    &          &      & (64, 1, 8)    &         \\ \hline
4 & Flatten            &    &          &      & 512           &         \\
  & Dense (drift rate) &    &          & 513  & 1             &         \\
  & Dense (boundary)    &    &          & 513  & 1             & Sigmoid
\end{tabular}
\end{table*}

\subsection{Decision SincNet Model}
\begin{figure*}[h]
\includegraphics[scale=0.25]{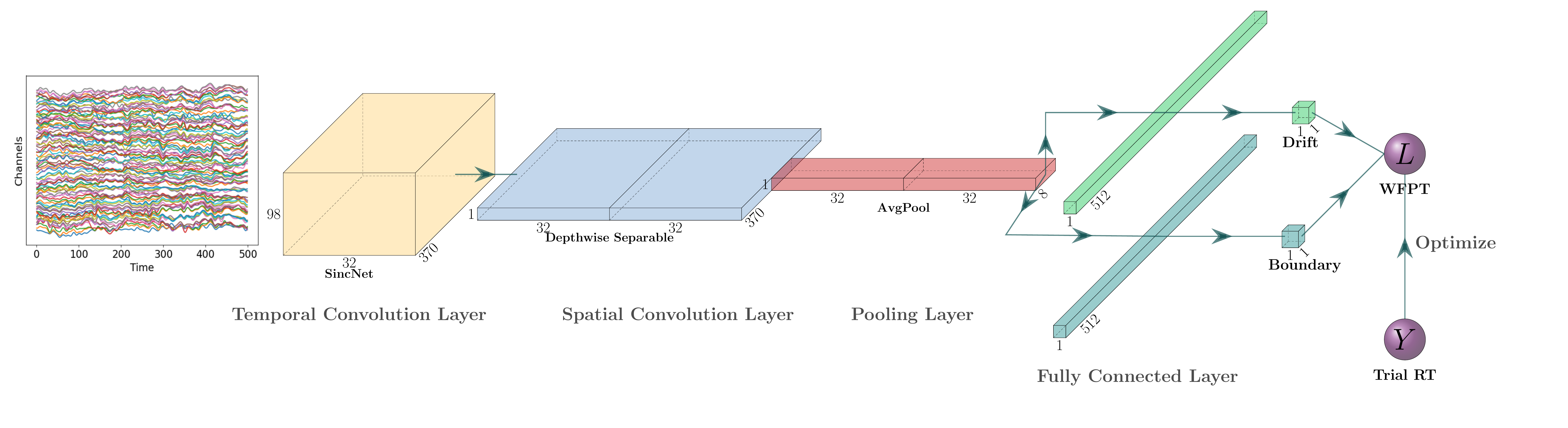}
\caption{Visualization of Decision SincNet model architecture.}
\label{fig0}
\end{figure*}

The Decision SincNet model is built to use trial-level EEG data to simultaneously predict drift rate and boundary separation in an individual participant. The model architecture is similar to Sinc-ShallowNet\cite{ravanelli2018speaker} and consists of four blocks, as shown in Table 1 and visualized in Fig.~\ref{fig0}. The model applies band-pass filters and spatial filters on the EEG data, pools the filtered data, and finally predicts the two parameters of the diffusion model in-parallel. There are 7556 trainable parameters in total. All the methods mentioned below may be reproduced by using the code on Github (https://github.com/jennyqsun/EEG-Decision-SincNet). The Deep Learning Models were developed in PyTorch \cite{paszke2019pytorch}, and trained from scratch using a workstation equipped with NVIDIA GeForce RTX 2080 Ti and 64 GB of RAM. On average, a model per subject takes five minutes to train.

The first block is a Sinc-convolutional layer\cite{borra2020interpretable}. Thirty-two kernels are parameterized by two cutoff frequencies of a Sinc Function. The low and high frequencies ($f_{1}$ and $f_{2}$) are trainable parameters of the model during learning, so the kernel can be expressed in the time domain as:

\begin{equation}
\boldsymbol{k}_j[n]=2f_2sinc(2\pi f_{2,j} n ) - 2f_1 sinc(2\pi f_{1,j}n)\label{eq}
\end{equation}
To avoid ripples in the signal, a Hamming window is applied to the bandpass filters\cite{ravanelli2018speaker}. 
Thus, the output from the SincNet layer is from a 2D Convolution between $i_{th}$ trial EEG data and the $j_{th}$ kernel $\boldsymbol{k}_j$. Since the kernel size has only one dimension, it is equivalent to 1D Convolution between signals from each channel $\boldsymbol{x}_{c}$ and $\boldsymbol{k}_j$:  

\begin{equation}
\boldsymbol{y}_{c,j}[n]= \boldsymbol{x}_c[n] * \boldsymbol{k}_j[n] = \sum_{\substack{l=0}}^{L-1} \boldsymbol{x}_c[n-l] \cdot \boldsymbol{k}_j[l] \label{eq}
\end{equation}
where $c \in [0,C-1]$,  $j \in [0,K-1]$ with $K$ representing the total number of kernels, $\boldsymbol{y}_{c,j}[n]$ is the filtered signals as the output from the Sinc layer, and $L$ is the length of the kernel. We set $L=131$ such that lowest central frequency possible is around 4 Hz when sampling rate is 500 Hz\cite{borra2020interpretable}.
    
The second block is a depthwise convolution layer where spatial kernels are applied to time series data with a depth of 2. Each temporally-filtered version of the input convolves with two spatial filters followed by the ReLU activation function. The third block consists of pooling operation. The window size and stride size are always set in a way such that they are equivalent to 250ms and 100ms, respectively, when scaling back to the original input. 

Following the Multi-task learning (MTL) paradigm \cite{ruder2017overview}, the last block has two "task-specific" layers that share all the previous hidden layers. MTL improves model performance by acting as a form of regularization, causing the model to prefer hypotheses that explain more than one task. The shared temporal- and spatial-filtered features are flattened and fed to two separate fully-connected (FC) layers to simultaneously produce two outputs, trial drift rate ($\delta$) and trial boundary ($\alpha$).
The two parameters are fit using the same loss function during training. Predicted drift rates are clamped such that $\hat{\delta} \in [-6,6]$, in order to avoid extreme values of the negative log likelihood.  A modified Sigmoid activation function, 
\begin{equation}
\Phi(z) = \exp \left( \frac{1}{1+\exp(-z)} \right)
\end{equation}
constrains the predicted alpha to a realistic range, $\hat{\alpha} \in (1,2.72)$. 

 To constrain the predicted value of boundary, output from its corresponding FC layer is followed by a modified Sigmoid activation, as shown in Fig.~\ref{fig2}:

\begin{figure}[ht]
\centerline{\includegraphics[scale=0.5]{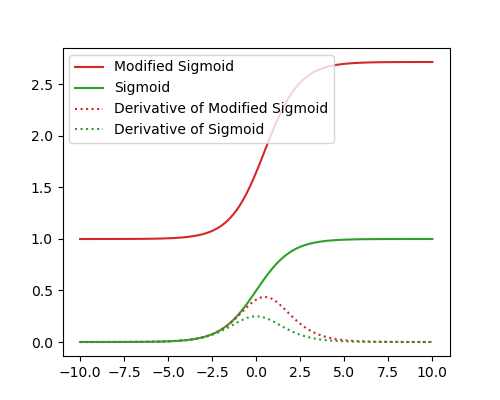}}
\caption{Demonstration of the modified Sigmoid activation function used for the FC layer to predict boundary.}
\label{fig2}
\end{figure}
The derivative of the modified Sigmoid is:

\begin{equation}\
\Phi'(z) =\frac{\exp({\frac{1}{\exp(-z)+1} + z)}}{(\exp(z)+1)^2}
\end{equation}

\subsection{Modeling the Diffusion Process}
Given a drift rate $\delta$, the DV could hit either of the boundaries due to noise. Variability across time in the diffusion process can drive the process to terminate at the wrong boundary by mistake. A diffusion model that views correct and incorrect choices as the two boundaries will produce two different shapes of probability distributions. Decision SincNet is designed to make a point estimate of diffusion model parameters using one EEG trial.  We modeled the two boundaries as the two choices, where drifting to the upper bound represents one choice, and drifting to the lower bound represents the process towards the other choice, and assumed there was no starting bias between the choices, which is a reasonable assumption for the data sets tested here. The response times for two choices are assumed to have identical probability density function and could be estimated using the same likelihood function. The main purpose of our approach is to directly predict model parameters from the EEG data on each trial. 

\subsection{Training}

Before training, each subject's data were split into 80\% for training and validation, and the other 20\% for testing. For each subject, 1 second of EEG trials after stimulus onset were used as inputs, and the corresponding RTs were used as training labels. Separate models are trained for each subject. Initial bandpass filters were randomly selected from a uniform distribution, $U\sim (1,32)$. 
Weights are updated by the gradient descent as
\begin{equation}
\boldsymbol{\theta}_{i} = \boldsymbol{\theta}_{i}  - \eta \frac{\partial }{\partial \boldsymbol{\theta}_{i}} \mathcal{L}(\boldsymbol{\theta})
\label{eq}
\end{equation}
$\eta$ is the learning rate and it was set to be $10^{-3}$. The model was trained using batch size of 64, and Adaptive moment estimation (Adam) optimizer was used. To avoid overfitting, early stopping technique was applied with a patience score of 20. 



\subsection{Model Evaluation}

\subsubsection {Trial-to-trial Variability of the Estimates} Spearman's Rank-Order Correlation was used to examine the direction and strength between estimated parameters and observed RTs. Strong correlations would suggest that the model captures trial-by-trial neural variability that are reflected in RTs and can be mapped onto drift rates and boundaries.

\subsubsection{Comparison Between Decision SincNet and Bayesian MCMC Models} The distribution of trial estimates for a subject obtained from Decision SincNet should fall within reasonable ranges. We compared our results to Bayesian MCMC model fits using JAGS sampler to estimate posterior distributions of the model parameters using only behavioral data. The fit of a basic model use the following 
prior structure\cite{nunez2017attention}:
\begin{equation}
\begin{split}
(\delta | \mu, \sigma) \sim \mathcal{N}(0,\,\sigma^{2}), \sigma^{2} \sim \Gamma(1,1) \\
(\tau | \mu, \sigma) \sim \mathcal{N}(0.5,\,\sigma^{2}), \sigma^{2} \sim \Gamma(0.3,1) \\
(\alpha | \mu, \sigma) \sim \mathcal{N}(1,\,\sigma^{2}), \sigma^{2} \sim \Gamma(1,1)
\end{split}  
\end{equation}
After obtaining the posterior distributions of the parameters, we compare the distribution of the trial estimates and check if they were within sensible ranges.


\subsubsection {Model Comparison by Log Likelihood of WFPT} 

To check whether EEG data have the resolution to give rise to meaningful trial-level estimates, we can compare the sum of negative log-likelihood $- \sum \log \text { Wiener }(t_i \mid \delta, \alpha, \beta, \tau) \triangleq \ell (\delta, \alpha \mid t)$ of the training RT between trial level estimates of parameters and median estimates of parameters. We compare the possible combinations of likelihood of single trial estimates with median estimates, namely, $\ell\left(
{\delta}, {\alpha}\mid t_i^{train} \right)$,
$\ell\left({\delta}, {\overline\alpha}\mid t_i^{train} \right)$,
$\ell\left({\delta}, {\overline\alpha} \mid t_i^{train} \right)$, 
$\ell\left(\overline{\delta}, \overline{\alpha}\mid  t_i^{train}\right)$. 
If the sum of the negative likelihood of either trial drift and trial alpha is smaller than that of median drift and alpha, we can conclude that the brain data is meaningfully linked to the trial to trial variability in behavior. 

\subsection{Generalization of the model}  To check whether the trained model could be generalized to unseen data, we compared sum of negative log-likelihood of test RT data using the median parameter estimates obtained from training data $\ell\left( 
\overline{\delta ^{train}}, \overline{\alpha ^{train}} \mid t_i^{test} \right)$  with all combinations of trial parameter estimates from trial EEG data $\ell\left({\delta ^{test}}, {\alpha ^{test}}\mid t_i^{test}\right)$ . Higher likelihood indicates that single trial EEG produced meaningful trial level predictions of DDM parameters.

\subsubsection{Predictive Uncertainty} Monte Carlo dropout \cite{gal2016dropout} is applied to estimate the uncertainty of predictive drift rates and boundaries. Mathematically, the use of dropout in a neural network can be viewed as Bayesian approximation of a Gaussian process \cite{damianou2013deep}. Dropout is enabled during predictions in which different neurons are randomly deactivated with a constant dropout rate. After the repeated random sampling process, we can approximate the following predictive distribution

\begin{equation}
p\left(\delta,\alpha \mid \boldsymbol{x}\right)=\int p\left(\delta, \alpha \mid \boldsymbol{x}, \boldsymbol{\theta}\right) p({\boldsymbol{\theta}}) \mathrm{d} {\boldsymbol{\theta}}
\end{equation}

\begin{figure}[h]
\centerline{\includegraphics[scale=0.25]{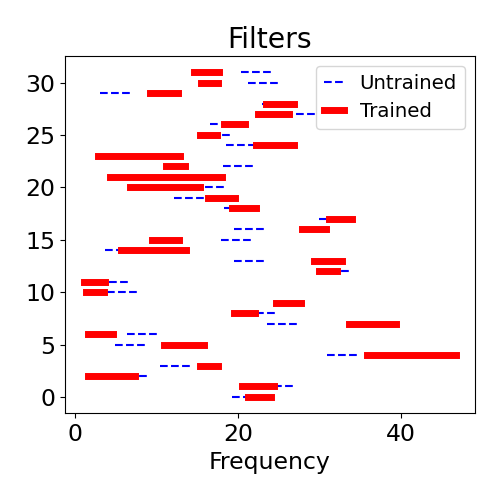}}
\caption{Comparison of filters. Dotted blue lines represent the random initialization of the bandpass filters before the model has been trained. Red lines represent the bandpass filters after the model has been trained }
\label{fig3}
\end{figure}

\begin{figure}[h]
\centerline{\includegraphics[scale=0.25]{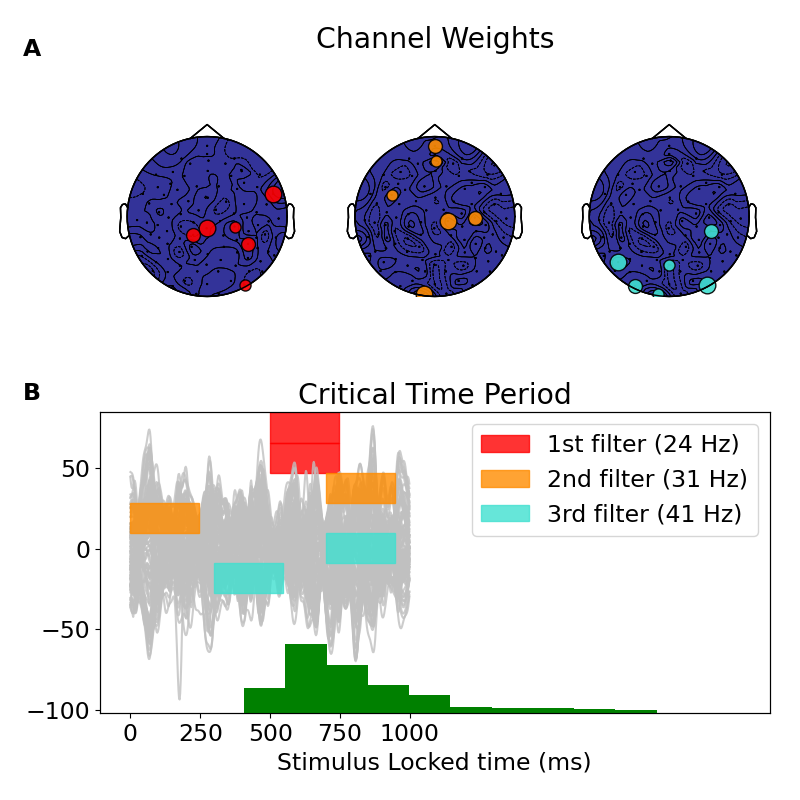}}
\caption{Post-hoc analysis of the trained model. Three colors represent the most critical frequencies band obtained from the normalized gradients. The corresponding central frequencies are given in parenthesis. Panel A shows the three most important pairs of weights from spatial filter. Weights are in pairs due to the depth of 2 in spatial convolution layer. Panel B shows the three time windows labeled by their corresponding frequencies that are most critical to predictions. }
\label{fig4}
\end{figure}

\begin{figure*}[!ht]
\centerline{\includegraphics[scale=0.28]{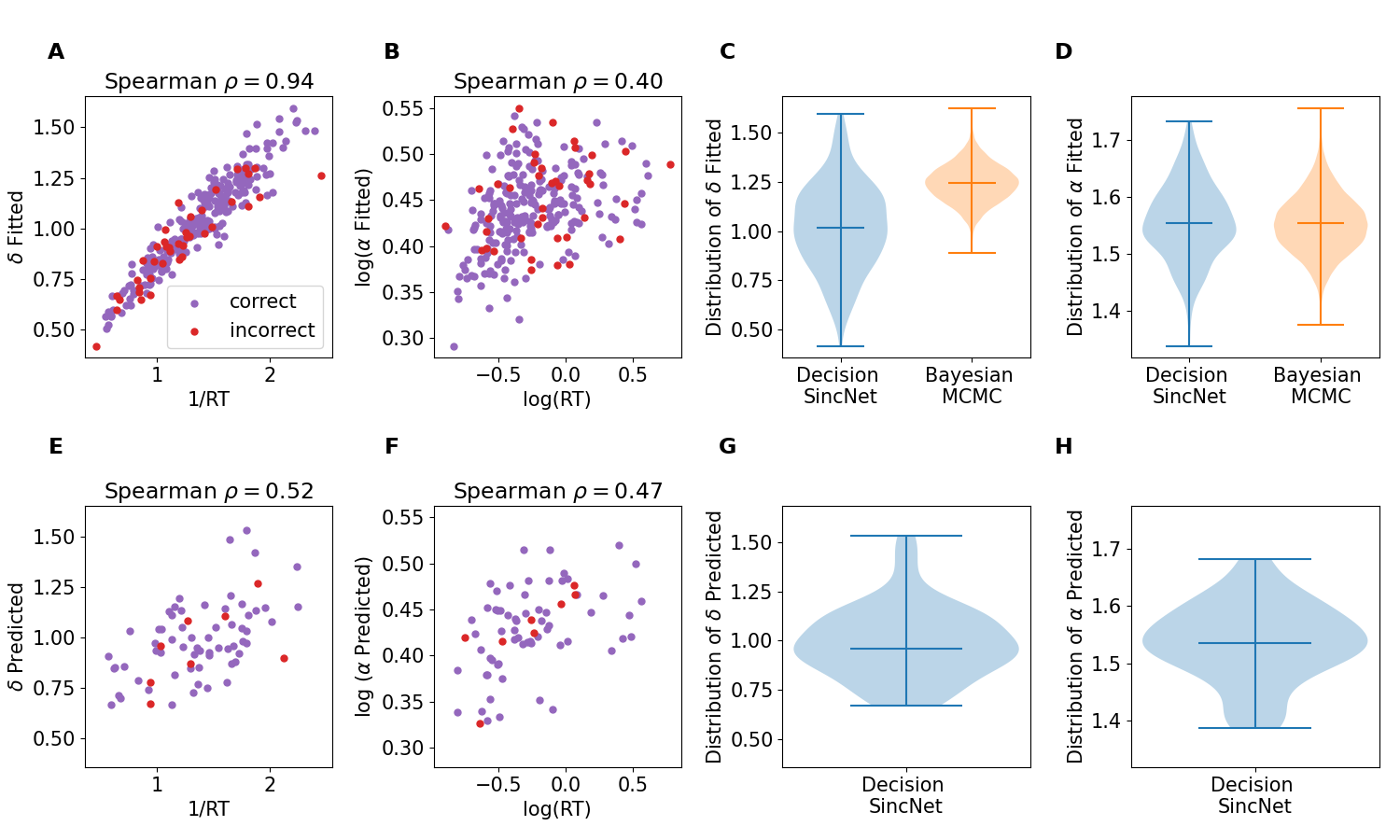}}

\caption{Model performance of one subject. First row are results of training data using the Decision SincNet model for this subject. Second row are results on test data. Panel A and B show model performance on fitted results. Scatter plots depict parameters predicted from trained model against observed RTs for each trial. Trials with correct responses  are labeled in purple, and trials with incorrect responses are labeled in red. Panel C and D, comparisons of distributions based on drift rates and boundaries estimated from Decision SincNet with Bayesian MCMC (without EEG data).  Distributions of Decision SincNet (blue) are plotted using all of the trial estimates, and distributions of Bayesian MCMC (orange) are plotted using the 30,000 samples obtained from MCMC when the Markov chains have converged. The three bars indicate the top, median, and bottom of the violin's distribution, respectively.  Panel E and F show model performance on test data. Panels G and H show distributions based on drift rates and boundaries predicted from test EEG data using Decision SincNet.  1/RT are used as a proxy of drift rates on x-axis in Panels A and E. Log(RT) and Log($\alpha$) is used in Panels B and F for visualization purposes.} 

\label{fig5}
\end{figure*}[b]

\subsubsection{Interpretability of Decision SincNet}
After training, we inspected the frequency bands that were learned. An example of interpretable bandpass filters from one subject is demonstrated in Fig.~\ref{fig3}. Each filter is represented by the cutoff frequencies $f_{1,j}$ and $f_{2,j}$. Filters learn to move from their initial random frequency bands and change the cutoff frequencies. We use gradient-based technique to interpret the learned bandpass filters and spatial filters for each subject following  \cite{borra2020interpretable}. We extend the method to further interpret pooled time windows. Critical temporal filters and time periods are evaluated based on the idea of saliency maps \cite{simonyan2013deep}. A forward pass is performed on an input $i$ of interest followed by the calculation of the gradient of an output $Y$ with respect to a feature of interest $F$ as $w=\left.\frac{\delta Y}{\delta F}\right|_{F_{i}}$. Important features are identified as the ones with the largest normalized gradient. The use of the gradient relies on the approximation of the output by applying a first-order Taylor expansion $Y(F) \approx w^{T} F+b$.  Fig.~\ref{fig4} shows an example from the same subject using the method. Three different color demonstrates three different frequency bands that are most critical to the prediction based on their normalized gradient. Absolute values of the spatial filters associated with a specific band-pass filter are then analyzed for the significance of each electrode. 
Panel A demonstrates the important weights identified. Each subplot shows two sets of important weights for a specific frequency band. This is because two separate spatial filters are applied to each filtered output from SincNet layer. Panel B demonstrates the important time windows obtained from the saliency map of the pooling layer, labeled by the center frequency of the corresponding critical frequency bands. In the example subject shown here, gamma band (31 Hz, 41 Hz) throughout the decision interval, and beta band (24 Hz) activity around the response were the most important bands.

\section{Results}
\begin{figure}[h]
\centerline{\includegraphics[scale=0.22]{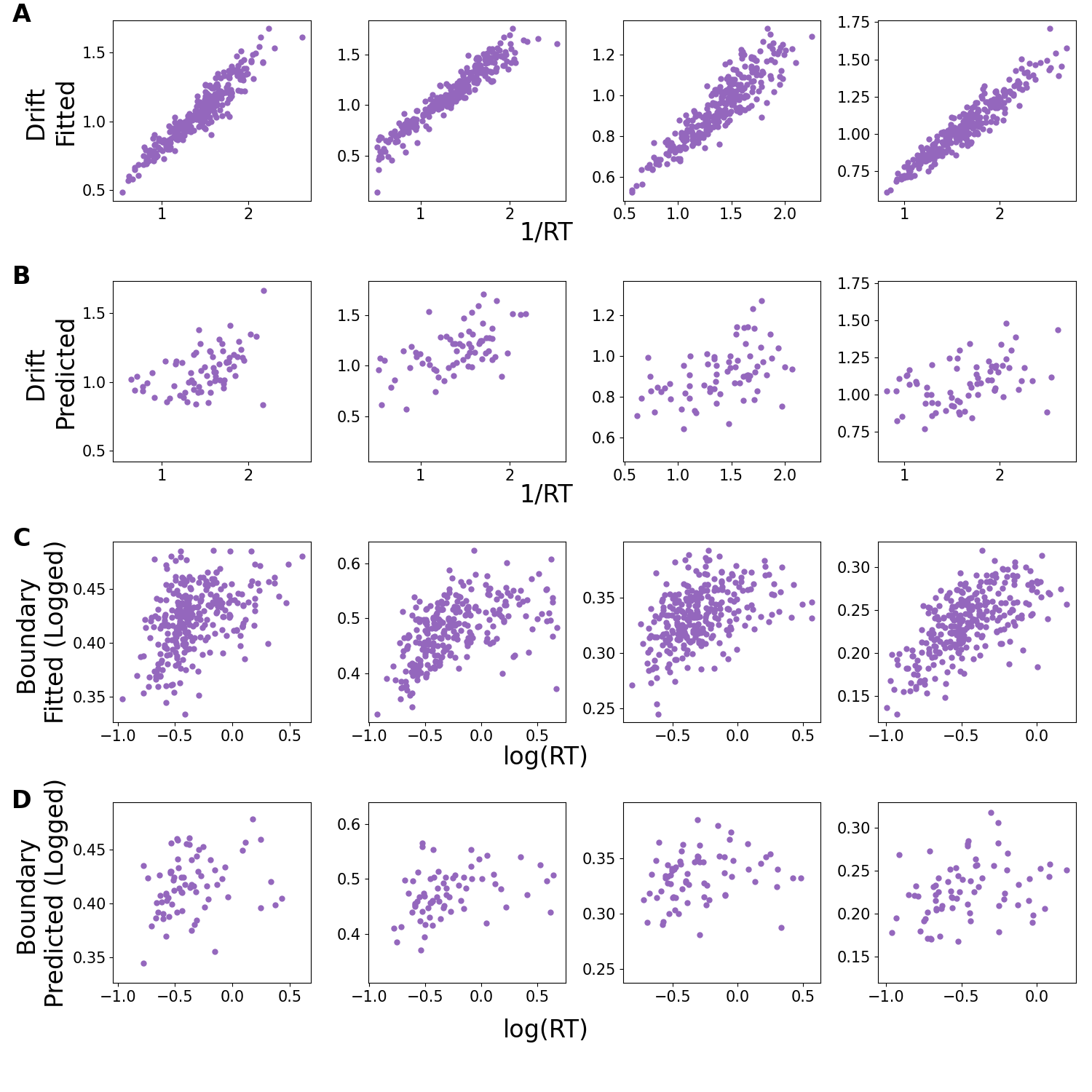}}
\caption{Scatterplots results of 4 example subjects. Each column represents the results of one subjects. Panel A shows the scatter plots of predicted drift against 1/RT with the training data. Spearman correlations from left to right are $\rho = 0.931^{***}, 0.963^{***}, 0.900^{***}, 0.940^{***}$. Panel B shows the scatter plots of predicted drift ($\delta$) against 1/RT with the test data. Spearman correlations are $\rho = 0.506^{***}, 0.540^{***}, 0.472^{***}, 0.429^{***}$. Panel C shows the scatter plots of predicted boundary log($\alpha$) against log(RT) with the training data. Spearman correlations are $\rho = 0.506^{***}, 0.426^{***}, 0.469^{***}, 0.631^{***}$. panel D shows the scatter plots of predicted log($\alpha$) against log(RT) with the test data. Spearman correlations are $\rho = 0.285^{*}, 0.391^{***}, 0.272^{*}, 0.336^{**}$. Correct and incorrect trials were both included. * Correlation is significance at the 0.05 level (2-tailed) **  Correlation is significance at the 0.01 level (2-tailed) ***Correlation is significance at the 0.0001 level (2-tailed)} 
\label{fig6}
\end{figure}

\begin{figure}[h]
\centerline{\includegraphics[scale=0.27]{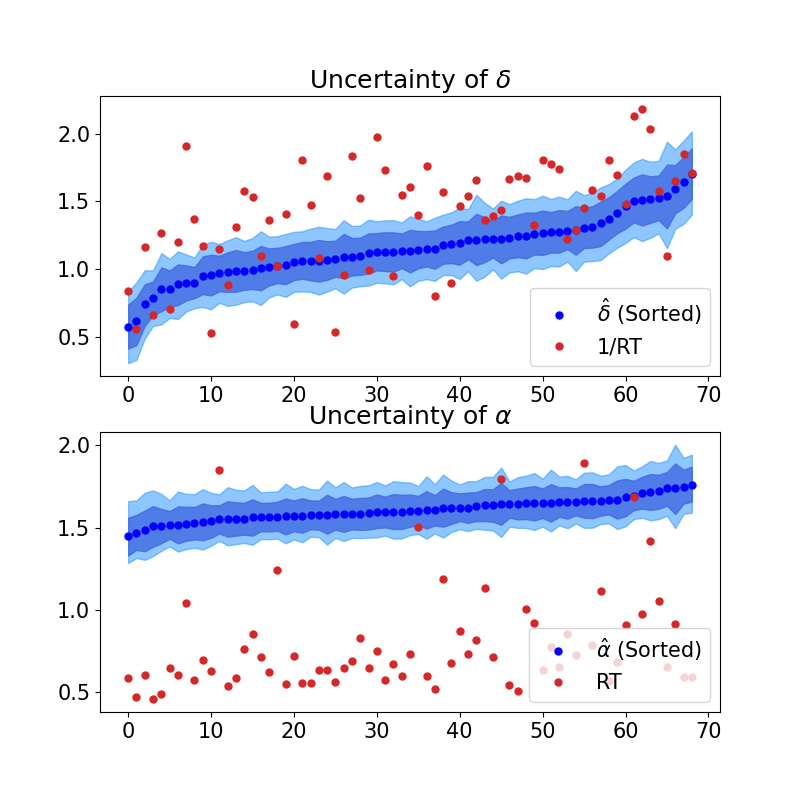}}
\caption{Uncertainty of the model prediction obtained from Monte Carlo Dropout. Blue dots in the top figure are median drift rate estimates sorted from small to large. Red dots are the corresponding 1/RT for that trial as proxy of speed. Dark blue area represents the Standard Deviation, and light blue area represents the 90\% Credible Interval. Blue dots in the bottom figure are median boundary estimates sorted from small to large. Red dots are the corresponding RT. Prediction were repeated 5000 times with different neurons deactivated.}
\label{fig7}
\end{figure}

Because Decision SincNet predicts two parameters that can not be observed directly, model performance was evaluated by focusing on the following issues: model comparison, trial-to-trial variability of the parameters within each subjects, model generalization to unseen data, and uncertainty of model prediction. 

Fig.~\ref{fig5} presents the comparison of model performance of one subject using the trained Decision SincNet model. Fig.~\ref{fig5}A and Fig.~\ref{fig5}B show the trial-by-trial correlations between estimated parameters and observed RTs from the training data.  The Spearman correlation between fitted drift rates ($\delta$) and 1/RT are strong ($\rho = 0.937, p <0.001$). 1/RT is used as a proxy of speed of information processing. Spearman Correlation between fitted alpha and RT are moderately strong ($\rho = 0.401, p <0.001$). Logarithm of RT and boundary is used to uncluster the data for visualization. 

Fig.~\ref{fig5}C and Fig.~\ref{fig5}D compare the distributions based on parameters estimated by different models from the training EEG data. Random samples from the posterior distributions based on Bayesian MCMC model are drawn to match to the samples size of Decision SincNet during training, and Kullback–Leibler (KL) Divergence was calculated for each of the parameters. Divergence of distribution of trial drift rates based on Decision SincNet from the posterior distribution of drift based on Bayesian MCMC is 77.714. Divergence of distribution of trial boundary based on Decision SincNet from the posterior distribution of alpha based on Bayesian MCMC is 0.347. These results suggest that single trial drift rates estimated from Decision SincNet contains more trial specific information, as it is more distinct from the posterior distribution of drift from the Bayesian MCMC. 

Fig.~\ref{fig5}E and Fig.~\ref{fig5}F show the trial-by-trial correlations between estimated parameters and observed RTs from test data, which the model has never been seen. The correlation between fitted drift rates and 1/RT are moderately strong ($\rho = 0.517, p <0.001$), and the correlation between fitted alpha and RT are moderately strong ($\rho = 0.468, p <0.001$). More examples examples are shown here in the Fig.~\ref{fig6}. Each column represents the performance of a subject, and each row represents the relationship between response times (x-axis) and parameters estimated from the model (y-axis). Spearman correlations are also reported in caption.  

We also compared the sum of negative likelihood as described above in Method Section (E.3). Our results indicate that using single trial estimates of both drift and boundary performs the best, as $\ell\left({\delta}, {\alpha}\mid t_i^{train}\right)$ has the lowest value. This suggests that the model was trained to extract sufficient brain data to distinguish RT variability on each trial by mapping brain signals to model parameters. When we performed out-of-sample prediction by using the trained model to predict unseen EEG data, all the likelihoods of the single trial estimates $\ell\left(
{\delta}, {\alpha}\mid t_i^{test}   \right)$,
$\ell\left({\delta}, {\overline\alpha}\mid t_i^{test} \right)$,
$\ell\left({\overline\delta}, {\alpha}\mid t_i^{test} \right)$, outperform the likelihood of the median estimates obtained from training data$,  \ell\left(\overline{\delta ^{train}}, \overline{\alpha ^{train}} \mid t_i^{test}  \right)$, implying a successful predictive model with ability to generalize to unseen data.

Fig.~\ref{fig7} is the results from Monte Carlo Dropout that captures the uncertainty of the trained model. We used the model trained for the subject and apply different dropout of neurons at test time for 5000 times. Blue dots in the top figure are median drift rate from the predictions sorted from small to large, and blue dots in the bottom figure are the median boundary estimates sorted. Red dots are the corresponding 1/RT for that trial as proxy of speed. Dark and light blue area represents the Standard Deviation, and the 90\% Credible Interval respectively. The posterior distributions of model predictions of both parameters fall within reasonable ranges. 

The model was able to fit and predict drift and boundary for all subjects (n=45) within reasonable ranges. During training, 41 out of 45 subjects showed the likelihood test of the single trial estimates of both parameters outperform the likelihood test of the median estimates, suggesting the benefit of single trial estimates for model fitting. When we test the trained model on unseen data, 14 out of 45 subjects have demonstrated that the likelihood of having either or both single trial estimates outperforms the likelihood of the median estimates, indicating that meaningful single-trial parameter estimates could be generalized to out-of-sample brain data for a subset of the subjects. We believe the performance in out-of-sample prediction would be improved with more data in each participant.

\section{Conclusion}

We proposed and validated the Decision SincNet, a  neurocognitive model of decision making that integrates EEG data and the drift diffusion model estimated by a neural network. The end-to-end model uses the likelihood function of Wiener first-passage time to train the model on one second of EEG data given the RT. We have shown that the model can simultaneously predict two cognitive parameters of Drift-Diffusion model that reflect speed and caution during decision making. The model uses interpretable convolutional layers, that automatically learned bandpass filters, channel weights, and critical time windows which can be inspected to understand EEG relationships to human cognition. Various methods of model evaluation and post-doc analysis of the model have also been discussed. Examples of the model results are reported.

Future work should aim to tackle some of the current challenges of this new approach. Even though single trial drift and boundary are often the best model during training, the average of fitted parameters sometimes remained the most useful estimates when testing the model on unseen data, suggesting low signal-to-noise ratio in single-trial EEG data in some participants. There are also cases where the model learns to map the trial-to-trial variability of only one of the parameters, but not both. To test whether this is a genuine observation, we need experimental data where each parameter is experimentally manipulated\cite{voss2004interpreting} with simultaneous EEG recordings. Although taking the average of boundary during optimization showed results that are more consistently with the results from Bayesian MCMC model, better optimization techniques should be developed to bypass the issue of overfitting boundary parameter. We have recently developed a follow-up approach to address this issue by additionally forcing the drift parameter to find solutions that correlate with the RT during optimization. We also believe the model performance could be improved by manipulating the Decision SincNet architecture, in order to fit the entire set of DDM parameters, including bias and non-decision time. Future work should also extend the model to incorporate likelihood functions for incorrect trials with a joint behavioral and EEG dataset that has lower accuracy. Finally, other explainable deep learning techniques for EEG signals such as occlusion sensitivity analysis for estimating source signals \cite{ieracitano2021novel} should be explored. 

\section*{Acknowledgment}

This research was supported by grants \#1850849 and \#2051186 from the United States (US) National Science Foundation (NSF). Special thanks to Keith Barrett and Mariel Tisby for help with EEG data collection, artifact removal, and additional analyses of this data.

\bibliographystyle{IEEEbib}
\bibliography{refs}

\end{document}